\documentclass[preprint]{aastex}
\usepackage{rotate}

\newcommand{\g}{\gamma}

\def\submitted#1{\gdef\@submitted{#1}}
\begin{document}

\title{Experimental Constraints on the Axion Dark Matter Halo Density}

\author{Stephen J. Asztalos, E. Daw, H. Peng, L. J Rosenberg and D.
  B.  Yu} \affil{Department of Physics and Laboratory for Nuclear
  Science, Massachusetts Institute of Technology, 77 Massachusetts
  Avenue, Cambridge, Massachusetts 02139}

\author{C. Hagmann\altaffilmark{}, D. Kinion\altaffilmark{}, W.
  Stoeffl,\altaffilmark{} and K. van Bibber\altaffilmark{}}
\affil{Lawrence Livermore National Laboratory, 7000 East Avenue,
  Livermore, California 94550 }

\author{J. LaVeigne, P. Sikivie, N. S. Sullivan, and D. B. Tanner }
\affil{Department of Physics, University of Florida, Gainesville,
  Florida 32611}

\author{F. Nezrick} \affil{Fermi National Accelerator Laboratory,
  Batavia, Illinois 60510-0500}

\author{D. M. Moltz} \affil{Lawrence Berkeley National Laboratory, 1
  Cyclotron Road, Berkeley, California 94720 }
\submitted{Draft version \today}
\begin{abstract}
  Most of the mass of the Milky Way galaxy is contributed by its halo,
  presumably in the form of non-interacting cold dark matter.  The
  axion is a compelling cold dark matter candidate.  We report results
  from a search which probes the local galactic halo axion density
  using the Sikivie RF cavity technique.  Candidates over the
  frequency range 550$\!$ $\le$ f $\le$$\!$ 810 MHz (2.3$\!$ $\mu$eV
  $\le$$\!$ m$_{a}$ $\le$ 3.4 $\mu$eV) were investigated.  The absence
  of a signal suggests that KSVZ axions contribute no more than 0.45
  GeV/cm$^3$ of mass density to the local dark matter halo over this
  mass range.
\end{abstract}
\newpage
\section{Introduction}  
Measurements such as the BOOMERANG and MAXIMA cosmic microwave
background radiation power spectra \citep{ber00,max00} and Type Ia
supernovae data \citep{per99,res98} suggest that dark matter makes up
the bulk of the matter content in the universe.  The agreement between
expectations from Big Bang nucleosynthesis \citep{bur99} and the
primordial abundance of the lightest elements \citep{osw00} strongly
constrains the total baryonic content to be a small value, implying
that the majority of the dark matter must be non-baryonic.  The
results of other diverse astrophysical measurements based on the
Sunyaev-Zel'dovich effect \citep{gre01}, strong gravitational lensing
\citep{yc00} and galactic flows \citep{dek99} lend powerful support to
this picture.

Although the nature of dark matter remains unknown, its gravitational
effect is pronounced: much of the dynamics of spiral galaxies cannot
be understood without there being a massive dark matter halo
\citep{sr00}.  Two categories of particle cold dark matter (CDM)
candidates have survived experimental and theoretical scrutiny over
time: the lightest supersymmetric particle \citep{g83,ell84} and the
axion \citep{wei78,wil78}.  Neutrinos and other forms of hot dark
matter are thought to contribute insignificantly to closure density
and, in any case, cannot explain structure formation.  Galaxy
formation requires CDM, i.e., dark matter which is already
non-relativistic at the time of decoupling.  Recently, experiments
have begun with the sensitivity either to detect or exclude possible
CDM halo candidates.  In this paper we present upper limits on the
local axion halo density derived from a search for cold dark matter
axions.
\section{Axion Physics}
The axion is the pseudo Nambu-Goldstone boson \citep{wei78,wil78}
associated with a new spontaneously broken global U$_{\rm{PQ}}$(1)
symmetry invented to suppress strong CP-violation \citep{pq77}. There
is some model dependence in assigning U$_{\rm{PQ}}$(1) charges to
particles: in the KSVZ scheme \citep{kim79,SVZ80} the axion only
couples to quarks at tree level, while in the GUT-inspired DFSZ model
\citep{z80,DFS81} it couples to both quarks and leptons. The axion
acquires a mass that scales inversely with the (unknown) energy scale
f$_{a}$ at which the U$_{\rm{PQ}}$(1) symmetry breaking occurs.
Initially f$_{a}$ was presumed to be the electroweak energy scale, but
such massive axions were quickly ruled out in, e.g., beam dump
experiments \citep{yas81}.  Subsequently, it was proposed that axions
possess such small couplings to matter and radiation that for all
practical purposes they would remain forever ``invisible''.  Shortly
thereafter, an experiment was proposed that could make even very light
axions detectable \citep{sik83}.  The U.S. axion search experiment is
predicated on this approach whereby the axion converts into a single
photon via the inverse Primakoff effect. We use a resonant cavity
permeated by a strong static magnetic field, where the large number
density of virtual photons from the field enhances axion decay.

The allowed axion mass is constrained to 10$^{-2}$ to 10$^{-6}$ eV.
Axions with a mass less than a few eV would have cooled the core of
supernova 1987a to such an extent that the distribution of neutrino
arrival times would be inconsistent with observation \citep{tur88}.
Even heavier axions have been ruled out by a variety of astrophysical
and terrestrial searches \citep{tur90, ggr90, res91, gne99}.
Conversely, if the axion mass is less than some value they would have
been overproduced in the early universe.  This lower mass limit has
been calculated for various axion production mechanisms under
different early-universe scenarios, e.g., ``vacuum realignment''
\citep{pww83, as83, df83}, ``string decay'' \citep{bs94, yky99, hcs01}
and ``wall decay'' \citep{chs99}.  The vacuum misalignment mechanism
provides a lower mass limit of \(\sim\) 10$^{-6}$ eV which we adopt
for our search strategy; the other mechanisms produce a value that is
in fairly close agreement.  Common to all of these mechanisms is the
misalignment of the axion field with respect to the CP-conserving
(minimum energy) direction when the axion mass turns on during the QCD
phase transition.  Axions produced in this way are very cold.  Their
typical momentum was of order the inverse of the horizon scale at the
QCD phase transition ($p_a \sim 10^{-8}$ eV)\ when the temperature was
of order 1 GeV. The cosmological energy density in these cold axions
is of order: \citep{kt90}
\begin{equation}
\Omega_a \sim  0.5({{\rm{\mu eV}}\over{m_a}})^{7/6}.
\end{equation}
Hence, if $m_a$ is of order a few $\mu$eV, the mass range where we
search, axions contribute significantly to the energy density of the
universe.  Studies of large scale structure formation support the view
that the dominant fraction of matter is in the form of CDM.  Since CDM
necessarily contributes to galactic halos by falling into the
gravitational wells of galaxies (halo axions in our galaxy possess a
virial velocity $\sim$ 10$^{-3}$c), there is excellent motivation to
search for axions as constituents of our galactic halo.

There is a substantial body of evidence that our own galaxy is
surrounded by a massive dark halo, though its exact properties are not
well-constrained \citep{z98, Al00}.  Of particular interest to this
paper is the local dark matter halo density, whose value depends on
the degree of halo flattening as well as the core radii of the various
dark matter components.  To derive a reliable mass density, one can
turn to parameterizations of the density distribution, rejecting
distributions which fail to match observational constraints, such as
reproducing the local rotation speed of 200-240 kmsec$^{-1}$
\citep{ggt95}. A key element of this approach is the use of
microlensing data to estimate the fraction of local dark matter that
is in the form of compact objects.  Employing this methodology, one
arrives at a halo density of
9.2$^{+3.8}_{-3.1}$$\times$10$^{-25}$gm/cm$^{3}$.  If massive compact
halo objects (MACHOs) comprise a negligible fraction of the local halo
density, then the above number is likely an underestimate.  The local
halo density may also be enhanced because of our proximity to a
possible dark matter caustic \citep{sk97}. Our experimental
analysis directly constrains the local density of the axionic
component of the halo, and as such is independent of astronomical
observations and assumptions.
\section{Experimental Technique}
\begin{figure}[h]
  \figurenum{1} 
\epsscale{1} \plotone{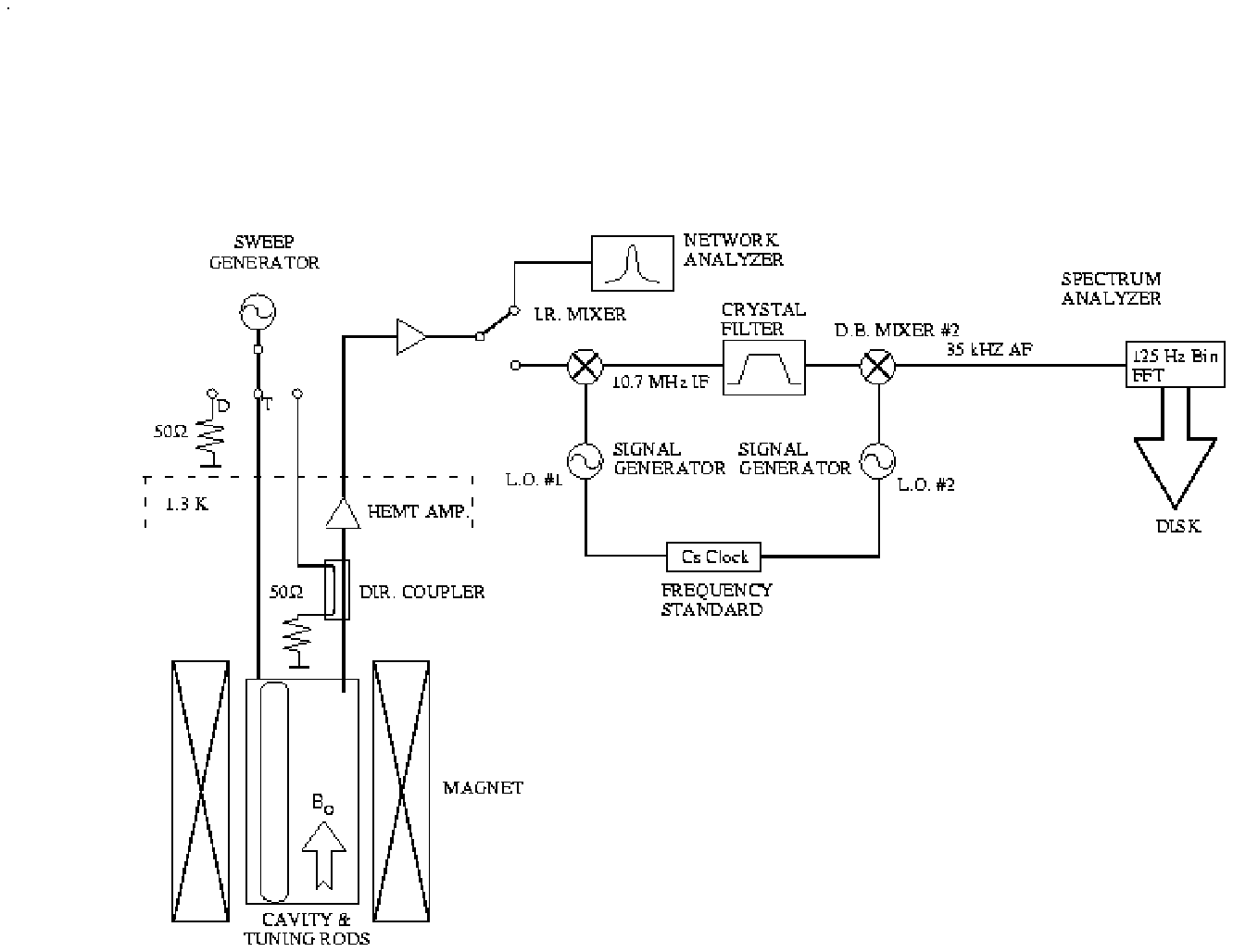}
  \figcaption[Schematic.eps]{The axion receiver chain. \label{fig1}}
\end{figure}

The interaction between axions and photons can be written as
\begin{equation}
\label{la}
\mathcal L = g_{a\g\g}a\vec{E}\cdot\vec{B},
\end{equation}
where \(g_{a\gamma\gamma}\) is the relevant coupling, \(a\) the axion
field and \(\vec{E}\) and \(\vec{B}\) the electric and magnetic
fields, respectively.  Since \(g_{a\g\g}\) is very small in the mass
range of interest, the spontaneous decay lifetime of an axion to two
real photons is vastly greater than the age of the universe.  In our
experiment, located at Lawrence Livermore National Laboratory, a
high-\(Q\) resonant cavity and superconducting magnet stimulate axion
conversion into a single real photon.  Resonant conversion occurs when
the cavity resonant frequency equals the axion rest mass.  Because
this mass is, a priori, unknown, resonant frequencies are changed by
moving either ceramic or metallic tuning rods from the wall to the
center of the cavity.  For a resonant cavity with a loaded quality
factor \(Q_L\), the axion-to-photon conversion power is
\begin{equation}
\label{pa}
P = 4 \cdot 10^{-26}\hbox{W}\left({V\over 0.22 \hbox{\scriptsize m}^3}\right)
\left({B_0\over 8.5\hbox{\ T}}\right)^2 C_{nl}\left({g_\gamma\over
0.97}\right)^2 \nonumber\\
\left({\rho_a\over{1\over 2} \cdot 10^{-24}
{{\scriptstyle g}\over\hbox{\scriptsize cm}^3}}\right) 
\left({m_a\over2\pi(\hbox{GHz})}\right)\hbox{min}(Q_L,Q_a),
\end{equation} 
where \(V\) is the cavity volume, \(B_0\) the magnetic field strength,
\(C_{nl}\)\ the mode-dependent cavity form factor, \(g_{\g}\) the
reduced coupling constant (equal to \(g_{a\g\g} \pi f_a/\alpha \)),
\(\rho_a\) the axion halo density and min\((Q_L,Q_a)\) the smaller of
either the cavity or axion quality factors. Typical values for the
first four parameters are 0.2 \(m^3\), 7.5 \(T\), 0.6 and 0.97,
respectively.  The copper cavity has a loaded (critically coupled)
\(Q_L\) $\sim$ 10$^{5}$, whereas \(Q_a\), the ratio of the
energy to the energy dispersion of the axion, is a factor of ten or so
larger over the present frequency range. The total power that results
from Eq.  \ref{pa} is of order $\sim$ $10^{-22}$ W; our cavity and
amplifiers are cooled to a few degrees Kelvin to minimize thermal
noise.  Fig.  \ref{fig1} is a schematic of the axion receiver chain.
A microwave signal centered at the cavity resonant frequency and
approximately 30 kHz wide is coupled out of the cavity by an electric
field probe and subsequently mixed down (in two stages) to near audio
frequencies.  At any given frequency 10$^{4}$ spectra are averaged by
fast-Fourier-transform (FFT) hardware, with each spectrum sampled for
8 ms.  The corresponding Nyquist resolution of 125 Hz is well matched
to the width ($\sim$ 750 Hz) of axions thermalized by interactions
with the galactic gravitational potential.

\begin{figure}[h]
  \figurenum{2} \epsscale{0.1}
  \includegraphics[angle=270]{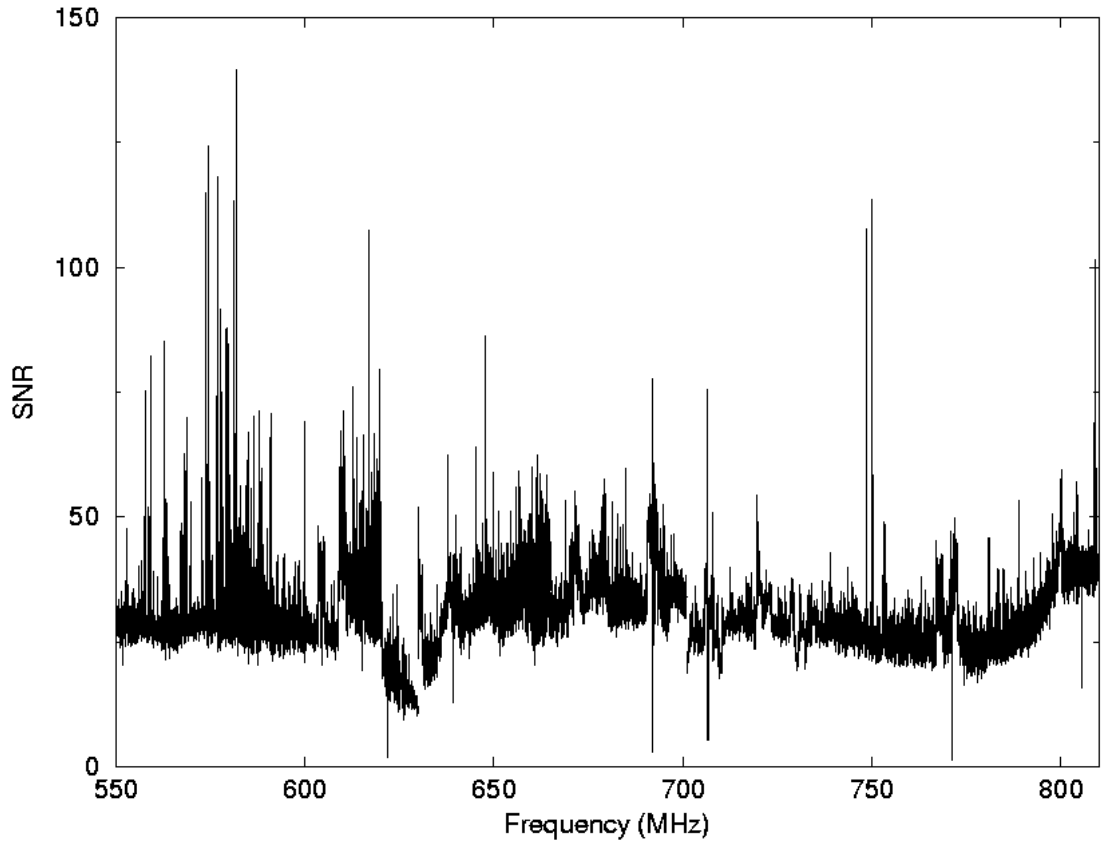}
  \figcaption[550-810_SNR.eps]{The SNR for the complete six bin
    combined data set.  The minimim target SNR ratio for these data is
    10. \label{fig2}}
\end{figure}

The Dicke radiometer equation \citep{dic46} dictates the integration time
necessary to achieve a specified signal-to-noise ratio (SNR)
\begin{equation}
\label{di} 
\rm{SNR}={P\over{T_n}}\sqrt{{t\over{B}}},
\end{equation}
where \(P\) is the axion power from Eq. \ref{pa} (times a factor which
accounts for the external coupling), \(T_n\) is the noise temperature,
\(t\) is the integration time and \(B\) is the axion bandwidth
(defined as \(f/Q_a\)).  Since the desired SNR is not attained in a
single pass over a given frequency interval, data in a single
frequency bin are the result of combined data from numerous
overlapping spectra.  The SNR for this multiple-pass data in the
interval 550 $\le$ f $\le$ 810 MHz is shown in Fig.  \ref{fig2}. A
second data set was formed by co-adding six neighboring bins into
single 750 Hz bins suitable for the virialized axion signal.  To
determine the number of candidate peaks that must be rescanned to
obtain an overall confidence limit of $\ge$90\%, artificial peaks are
injected into the data via software.  As the cut threshold is lowered,
the number of candidate peaks increases rapidly.  A typical cut
threshold of 2.3 $\sigma$ (6-bin) applied to our data yields numerous
candidates which are rescanned to a SNR commensurate with the original
data.  These data are subsequently added to the original data and from
these combined data sets a reduced set of candidates is generated and
scanned at the corresponding frequencies.  A final round of
data-combining produces a persistent-candidate list.  Candidates above
a threshold of 3.5 $\sigma$ in these data are manually inspected.  A
detailed description of the experiment and analyses may be found in
\citep{NIM00,asz01}.
\section{Results}
\begin{figure}[h]
  \figurenum{3} \epsscale{1.0} \plotone{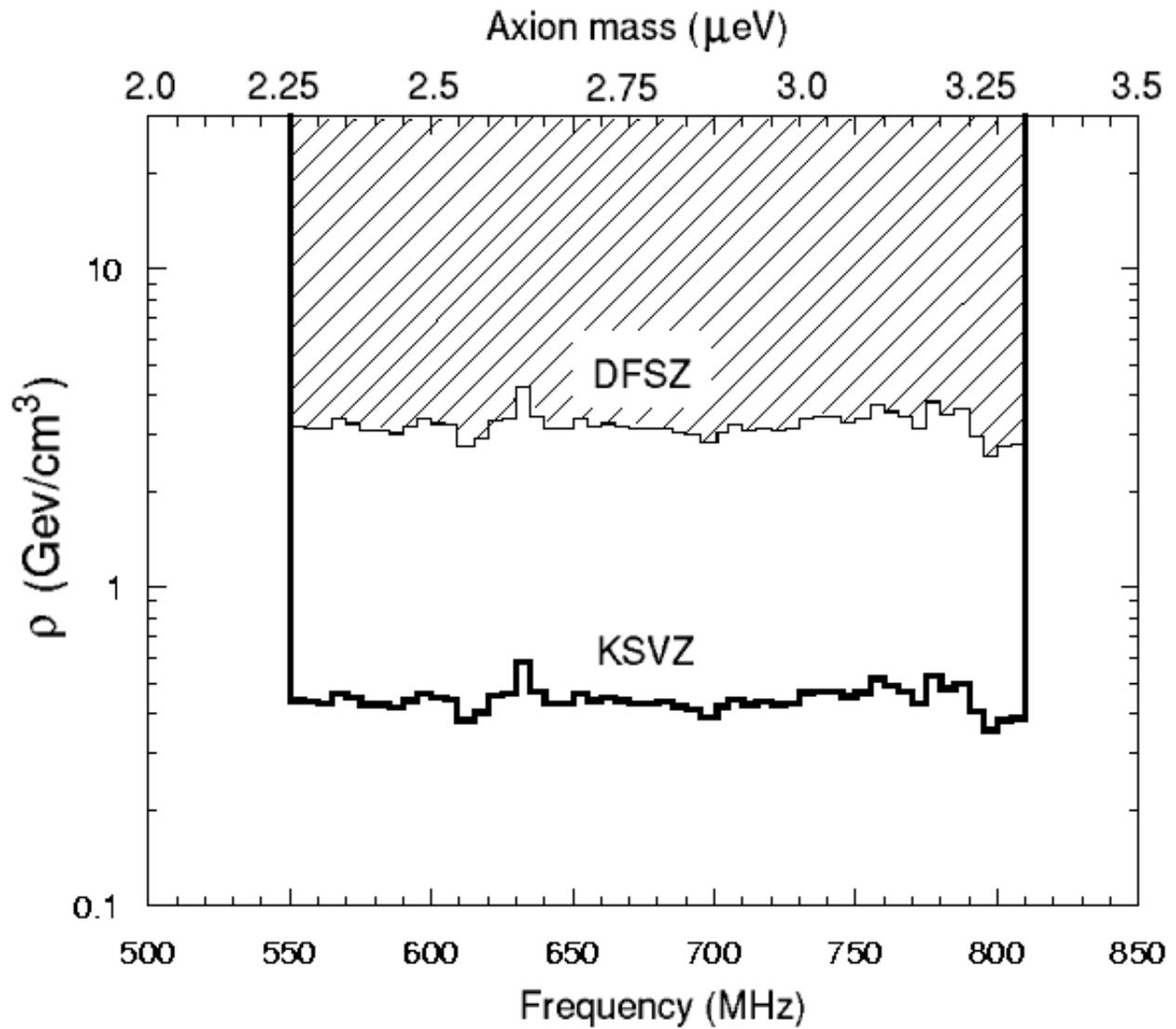}
  \figcaption[halodensplot24may01.ps ]{Excluded axion dark matter halo
    densities as a function of mass and frequency for both KSVZ (lower
    curve) and DFSZ (upper curve) axions. \label{fig3}}
\end{figure}

We have examined these data for candidates in each of
2.08$\times$10$^{6}$ 125 Hz and 750 Hz bins in the region 550 $\le$ f
$\le$ 810 MHz. A total of 13712, 1369 and 34 candidates survived each
stage of 6-bin data cuts, respectively.  All 34 persistent candidates
have been identified with strong external radio peaks.  To derive an
upper limit on the axion contribution to the local halo density, we
fix the axion-to-photon coupling \(g_{a\g\g}\) at the KSVZ level and
invert Eq.  \ref{pa} to calculate \(\rho_a\) as a function power
deposited in the cavity and axion mass.  The absence of a persistent
signal in these data over this range permits us to impose the limits
shown in Fig.  \ref{fig3}, where we plot the excluded axion dark
matter halo densities for both KSVZ (lower curve) and DFSZ (upper
curve) axions as a function of axion mass and frequency over the
interval 550 $\le$ f $\le$ 810 MHz.  The small variations in these
density limits represent effective integration times somewhat longer
or shorter than that prescribed by Eq.  \ref{di}.  The nominal
excluded mass density lies near 0.45 GeV/cm$^3$ for KSVZ axions and
3.0 GeV/cm$^3$ for DFSZ axions.  The former is comparable to the best
estimate of the local dark matter halo density.
\section{Conclusions}
There is abundant evidence that our own galaxy, like other spiral
galaxies, contains a vast dark matter halo.  Observation can neither
differentiate the various candidates, nor well constrain other
parameters that describe the halo, e.g., the local dark matter
density.  Since 1995 we have been using a single resonant cavity to
search for axions which may constitute the local dark matter halo over
the frequency interval 550 $\le$ f $\le$ 810 MHz.  The lack of a
persistent signal allows us to exclude the axion from contributing
more than 0.45 GeV/cm$^3$ to the halo dark matter mass density over
the mass range of 2.3$\times $10$^{-6}$ $\le$ $m_a$ $\le$ 3.4$\times
$10$^{-6}$ eV, should axions couple only to hadrons according to the
KSVZ prescription, with 90\% confidence.  This restriction is relaxed
to around 3.0 GeV/cm$^3$ in the DFSZ model, also with 90\% confidence.
It should be noted that other KSVZ- and DFSZ-like implementations
exist.  Some of these models (including some DFSZ-like models) give
rise to coupling constants that are larger than the benchmark KSVZ
\(g_{a\g\g}\) used in this paper \citep{kim98}.  These, too, are ruled
out by our results over the mass range quoted above.
\section{Acknowledgments}  This work was supported by the U.S.
Department of Energy under Contract Nos. DE-FC02-94ER40818, W-7405-
ENG-48, DE-FG02-97ER41029, DE-AC02-76CH0300, DE-FG02-90ER40560,
DE-AC-03-76SF00098, and the National Science Foundation under Award
No. PHY-9501959.

\newpage

\end{document}